\documentclass[aps,prb,twocolumn,showpacs]{revtex4}

\usepackage{graphicx}
\DeclareGraphicsExtensions{.png,.jpg,.eps}

\usepackage{xcolor}

\begin{document}

\title{Quantum Hall effect in gapped graphene  heterojunctions }

\author{ J. L. Lado$^{1,2}$, J. W. Gonz\'alez$^1$, and J. Fern\'andez-Rossier$^{1}
\footnote{On leave from Departamento de F{\'i}sica Aplicada, Universidad de Alicante,  Spain}$}
\affiliation{
(1)  International Iberian Nanotechnology Laboratory - INL,
Av. Mestre Jos\'e Veiga, 4715-330 Braga, Portugal
\\
(2) Departamento de F{\'i}sica Aplicada, Universidad de Santiago, Santiago de Compostela, Spain
}

\date{\today} 

\begin{abstract} 
We model the quantum Hall effect in heterostructures  made of  two gapped graphene stripes  with different gaps, $\Delta_1$ and $\Delta_2$.
We consider two main situations, $\Delta_1=0,\Delta_2\neq0$ and $\Delta_1=-\Delta_2$.   They are different in a fundamental aspect: only the latter feature  kink states that, when intervalley coupling is absent, are protected against backscattering. We compute the two terminal conductance of   heterostructures with  channel length  up to $430$ nm,  in  two transport configurations, parallel  and perpendicular  to the interface. 
By studying the effect of disorder on the transport along the boundary, we quantify the robustness of kink states with respect to backscattering. 
Transport perpendicular to the boundary shows how interface states open  a backscattering channel for the conducting edge states, spoiling the perfect conductance quantization featured by the homogeneously gapped graphene Hall bars. Our results can be relevant for the study of graphene deposited on hexagonal Boron-Nitride as well as to model graphene with an interaction-driven gapped phase with two equivalent phases separated by a domain wall. 
\end{abstract}
\maketitle

\section{Introduction}

The Hall conductance in quantum Hall bar is so accurately described by $\sigma_{xy}=n \frac{e^2}{h}$, where $n$ is an integer number, that
it is used\cite{VK} as our standard definition of the ratio of such fundamental constants as the square of the electron charge $e^2$ and the Planck constant $h$.
The origin of this extraordinary quantization, by which the conduction properties of a device are independent of the material properties,   is intimately linked to the fact that in quantum Hall bars transport takes place  only through the edges which host chiral states for which backscattering is  forbidden.\cite{Laughlin81,Halperin82}  In turn, the existence of chiral edge states that permit non-dissipative transport is warranted by  the topological order of the electronic states of the two dimensional gas  states.   For two decades, this state of affairs was observed at cryogenic temperatures, under high applied magnetic fields,  in two dimensional electron gases, hosted by carefully designed modulation doped  semiconductor heterostructures.  The discovery of quantum Hall effect on graphene\cite{Geim05,Kim05} 
even at room-temperature\cite{Novoselov_RT} on one side, and the proposal\cite{Kane-Mele1,Kane-Mele2,Zhang06} and subsequent discovery\cite{Konig2007} of quantum spin Hall insulators on the other, have dramatically expanded the materials and experimental conditions under which non-dissipative quantum transport linked to topological order can occur.

Most of the striking electronic properties of graphene are related to the absence of a gap separating the conduction and valence bands  which can thereby be described in terms of massless Dirac fermions.\cite{Semenoff84,RMP07} In particular, the magneto-electronic properties of graphene are fundamentally different from a non-relativistic two dimensional electron gas on three counts:\cite{Gusynin05}  the existence of  two identical sets of Landau levels,  for electrons and holes, the scaling of their energy with $\sqrt{B}$, as opposed to linear scaling of non-relativistic fermions, and  the existence of the $n=0$ Landau level with  zero energy.  These properties make the quantum Hall effect in graphene\cite{Geim05,Kim05} different from the one originally discovered in GaAs two dimensional electron gases.\cite{VK}  

There are several physical scenarios that motivate the study of the electronic properties of gapped graphene. First,  a gap could be opened by interaction driven electronic order,\cite{Min2008A,Araki2011,Semenoff2011} specially when a high magnetic field is applied. Second, as a result of the influence of the substrate like SiC\cite{Zhou2007,SIC-review} or  hexagonal Boron Nitride (BN),\cite{giovanetti2007}  although the lattice mismatch is known to complicate  this second possibility.\cite{Hone2010,Xue2011,Kindermann2012,Hunt2013}   
Third, BN itself can be described with the tight-binding Hamiltonian of gapped graphene and  the $\vec{k}\cdot\vec{p}$ Hamiltonian of other two dimensional materials with hexagonal symmetry, such as MoS$_2$, can be described with a massive Dirac Hamiltonian.\cite{Xiao12,Li12,Kormanyos} 
Fourth, intrinsic spin orbit coupling also opens a gap in graphene,\cite{Kane-Mele1,Kane-Mele2} with different sign at the two valleys, albeit very small.\cite{Yao07} 

  All of this  leads to the question of how  magnetotransport  properties of graphene and graphene-like materials change when a gap opens or, in  the long wavelength limit,   how massive and massless Dirac fermions are different in their reaction to an applied magnetic field. 
It turns out that, when the gap is opened by a constant staggered potential,  i.e., a potential that acts with opposite sign in the two sublattices of graphene,  as it happens if pseudospin magnetism, or in the case of hexagonal Boron Nitride, the answer to the question  is quite straightforward from the theory standpoint. This occurs because there is a simple  one-to-one relation between the energy levels and wave functions of a bipartite lattice Hamiltonian with no staggered potential and those of the same lattice when a constant staggered potential  is added.\cite{Pereira:2008,Soriano-JFR2012}  This relation permits to anticipate that   quantum Hall effect of massive Dirac fermions is  much closer to the one of massless Dirac fermions than to the one of non-relativistic electrons.\cite{Vasilopoulos2012}

\begin{figure}[t!]
\includegraphics[clip,width=0.5\textwidth,angle=0]{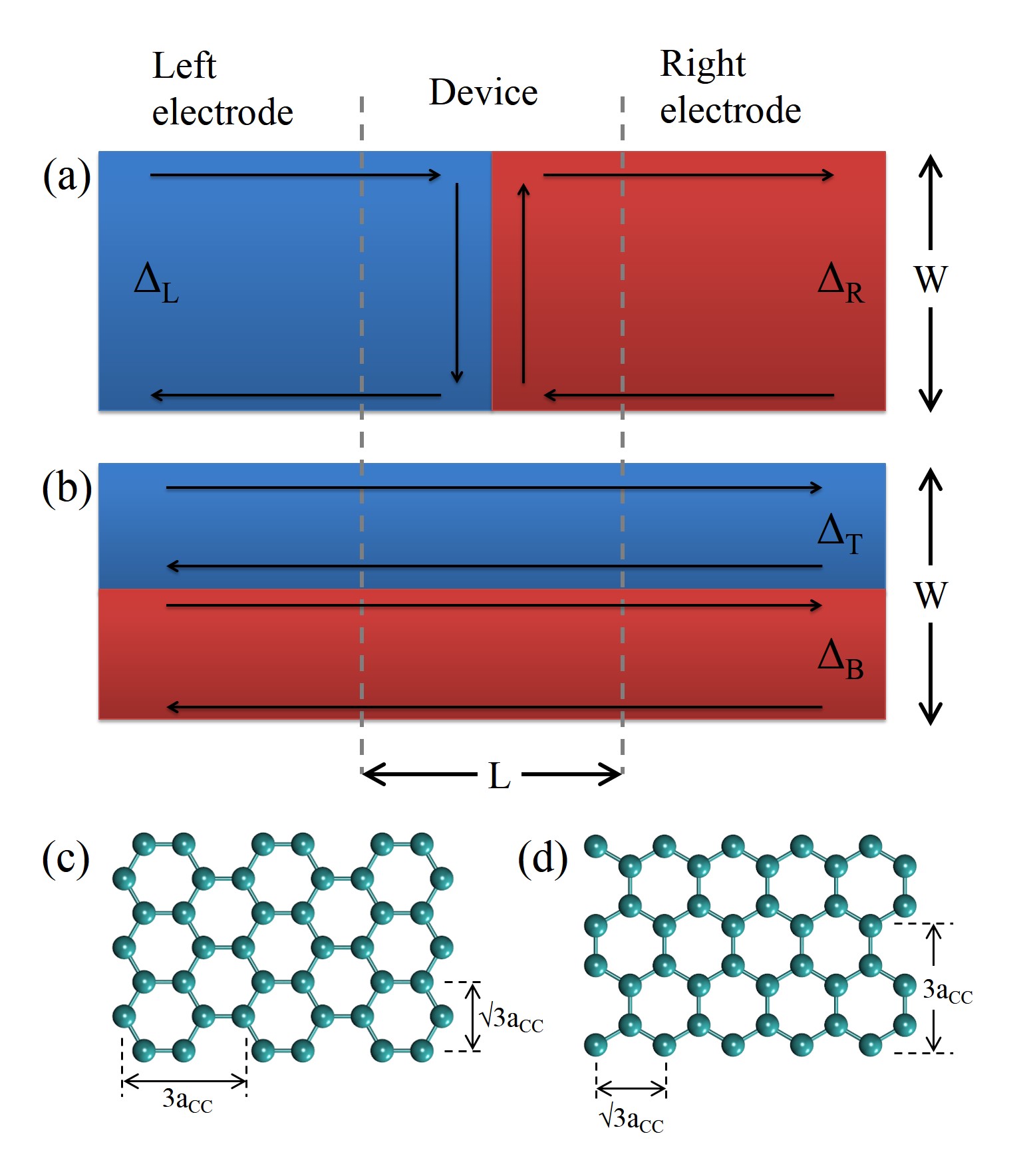} 
\caption{(Color online) Scheme of the two geometries considered in the text:
(a) perpendicular transport to the interface between the two different gap regions and (b) for the parallel transport.
Arrows indicate  edge, interface and kink states in the quantum Hall regime. 
(c) Detail of the armchair ribbon $N_W = 7$ $ (W=0.74$ nm$)$ and   $N_L = 3$  $(L=1.3$ nm$)$ and (d) zigzag ribbon edges $N_W = 4$ $(W=0.9$ nm$)$ and $N_L = 5$  $(L=1.2$ nm$)$.}
\label{Fig1:band}
 \end{figure}

The situation becomes more interesting when the gap -or mass-, is not 
homogeneous.  This could be the case, for instance,  of
 a heterostructure made of  two graphene or graphene-like materials with different gaps 
  $\Delta_1$ and $\Delta_2$, such 
 as the atomic layers of hybridized BN and graphene domains,\cite{Ci-NatMat10,Jung_NL} or if the gap is substrate-driven and, due to lattice mismatch,  features amplitude modulations larger the graphene unit cell.\cite{Kindermann2012,Hunt2013}
  Here we study the electronic properties of heterojunctions formed 
  between two gapped graphene-like systems, with different gaps, in the quantum Hall regime.  
  When decoupled, both Hall bars would have its own set of chiral edge states.   When coupled, the way edge states
  merge  determines the electronic properties of the interface, which is the  focus of this work.  

As we discuss below, we encounter two types of interface states.   The hybridization of 
pre-existing edge states that carry electrons in opposite directions will give rise to   interface states that can carry electrons in both directions.   In contrast,  the merger of two bars with opposite gaps gives rise to two interface states whose energy lies within the gap. At zero magnetic field these states can be rationalized as follows. When restricted to one  valley $\tau$,  graphene  electrons can be assigned a Chern number $\tau\frac{|\Delta|}{2\Delta}$, with $\tau=\pm 1$. According to the  index theorem\cite{Volovikbook}, the interface between two insulators with Chern numbers $n_1$ and $n_2$ should host ${\cal Q}=\left| n_1-n_2\right|$  domain wall states, which will correspond to zero modes  in the case of one dimension \cite{Jackiw-Rebbi} and chiral bands or kink states in two dimensions. 
 These are very similar to the recently discussed kink states in the interface of two graphene bilayers\cite{Martin-Blanter-Morpurgo,Li-MBM,Alvaro-APL} and multilayers\cite{JeilJung2011} with a gap opened by the application of an electric field with opposite direction at the two sides of the junction.

The rest of this paper  is organized as follows. In section II we 
review the electronic structure of gapped graphene under the influence of a 
perpendicular magnetic field, both within the conventional tight-binding model 
  as well as its long wavelength counterpart, 
the massive Dirac fermions. In section III we study the electronic structure 
of graphene heterojunctions in a geometry that preserves translational invariance along one direction,  which simplifies the discussion and  permits to unveil 
the appearance of  interface states.  In section IV we study the  quantum transport properties of the kink  states 
in  these structures, including the effect of disorder. In section V we consider heterojunctions formed by two semi-infinite Hall bars made of gapped graphene-like systems with different gaps.  
Transport in this type of heterojunction could be used to probe the interface states, which   enables backscattering between the otherwise decoupled chiral edges states.  In section VI we summarize our conclusions. 

\section{Stripes of gapped graphene in a magnetic field}
\subsection{Tight Binding Hamiltonian}
Here we review the well studied tight-binding model for graphene under the influence of a perpendicular magnetic field.\cite{RMP07,Gusynin05,Sigrist2000,Wakabayashi2001}
 We consider the standard tight-binding model for graphene, with one orbital per atom and first neighbor hopping t $\approx 2.7$ eV, with a staggered potential  $\Delta(\vec{R})$ that might be position dependent.  
 A given  graphene unit cell, located at $\vec{R}$, has two  atoms, denoted by  $A$ and $B$. Using this notation, the zero field tight-binding Hamiltonian reads:
  \begin{eqnarray}
 H=\sum_{\vec R} \left(  a^{\dagger}_{\vec{R}},b^{\dagger}_{\vec{R}}\right)
 \left(\begin{array}{cc} \Delta(\vec{R}) \delta_{\vec{R},\vec{R}'} &  t_{\vec{R},\vec{R}'} \\
 t_{\vec{R},\vec{R}'}& - \Delta(\vec{R})\delta_{\vec{R},\vec{R}'}  \end{array}\right)
 \left(\begin{array}{c}
 a_{\vec{R}'}
  \\
 b_{\vec{R}'}
  \end{array}\right),
  \end{eqnarray} 
  where $t_{\vec{R},\vec{R}'}$ is nonzero only for first neighbors and $a_{\vec{R}}, b_{\vec{R}}$ annihilate an
  electron at the $A$ and $B$ sites of unit cell $\vec{R}$ defined on a honeycomb lattice, and taking   $\Delta$ as a constant along the entire system,  this Hamiltonian describes graphene with a gap of $2 \Delta$ in both valleys.
  In the rest of this paper the spin degree of freedom is ignored.  
  Results for non-interacting   electrons with spin can be obtained by  adding the Zeeman shift to the obtained bands.  
 
Within this tight-binding description,  the effect of the applied magnetic field 
 is   included  replacing the hopping $t_{1,2}$  between sites $1$ and $2$ of the lattice of the $B=0$ Hamiltonian by  
  $t_{1,2} \rightarrow t_{1,2} e^{i\Phi_{1,2}}$
where\cite{Peierls,Sigrist2000,Wakabayashi2001} 
\begin{equation}
\Phi_{1,2}= \frac{e}{\hbar} \int_1^2 \vec{A}\cdot d\vec{r},
\label{peierls_phase}
\end{equation}
is the circulation of the vector potential $ \vec{A}$ associated to the magnetic field $\vec{B}$ and the labels $1$ and $2$ stand for the coordinates of the two atoms whose hopping integral is being calculated.  
 This is the lattice analogous of the canonical substitution for the free electrons,
 where the momentum operator $\vec{p}$ is replaced by $\vec{p}-e\vec{A}$. 
 Notice that the phase $\Phi_{1,2}$  that modulates the hopping is proportional to the ratio of the magnetic flux per unit cell and the magnetic flux quantum $\Phi_0=\frac{h}{e}$. 

In the following 
we assume that graphene lies in the $z=0$ plane,  and we take $\vec{B}=B(0,0,1)$.  Taking advantage of the gauge symmetry, we choose 
 \begin{equation}
 \vec{A}=B (-y,0,0),
 \label{gauge}
 \end{equation}
   so that 
\begin{equation}
 \Phi_{1,2}=-\frac{e B}{\hbar}\frac{(x_2-x_1)(y_2+y_1)}{2},
\end{equation}
where $(x_i,y_i)$ are the cartesian coordinates to atoms $1$ and $2$.  With this choice,  the Hamiltonian keeps translational invariance along the $x$ direction.

\subsection{Effective mass approximation}
  Whereas the tight-binding approach provides a fairly complete description of the non-interacting 
  electrons in graphene under the effect of a magnetic field, as we discuss below,  most of the results 
  for states with energies in the neighborhood
  of the Dirac points  can be rationalized 
by  making use of the $\vec{k}\cdot\vec{p}$ description of the bands
in the continuum limit.
\cite{Khon-Luttinger} 
  The magnetic field introduces a new length scale in the problem: 
  \begin{equation}
  l_B=\sqrt{\frac{\hbar}{eB}}.
  \label{mag_scale}
  \end{equation}
   We  assume a  sufficiently high  magnetic field so that $l_B<W$, 
  where W is the width of the ribbon, and hence the bulk quantum states become localized, the spectrum of states away from the edges becomes discrete,  the bulk is an insulator,  and dispersive and conducting states are only possible at edges.
  For typical magnetic fields, we also have  $a<<l_B$, 
  where $a=\sqrt{3}a_{CC} $ is the graphene lattice constant, which 
enable a  description of the energy levels in terms of an effective $\vec{k}\cdot\vec{p}$ Hamiltonian.  
  
   The effective   $\vec{k}\cdot\vec{p}$ or effective mass Hamiltonian turns out to be  isomorphic to the Dirac Hamiltonian at the two valleys:\cite{Semenoff84}
\begin{equation}
H_{\tau} = v_F \left( \Pi_x \sigma_x + \tau \Pi_y \sigma_y\right)  + \Delta\sigma_z,
\end{equation}
where $\vec{\Pi}\equiv \vec{p}- e\vec{A}$ is the canonical momentum operator,  
$v_F=3 t a_{CC}/2 \hbar$,  
$\vec{\sigma}$ are the Pauli matrices describing the graphene sublattice degree of freedom and $\tau=\pm 1$ describes the valley index.

Using the  gauge defined in Eq.  (\ref{gauge}) leads to
 \begin{eqnarray}
 H_{\tau}=
\left(
\begin{array}{cc}
\Delta & v_F \left [ p_x-eBy+i \tau p_y \right ] \\
v_F\left [p_x-eBy-i \tau p_y \right ] & -\Delta \\
\end{array}
\right), 
\label{HAMIL-KP}
\end{eqnarray}
This Hamiltonian is translationally invariant along the $x$ direction, so that we can assume its eigenfunctions are products $e^{ik_x x}\vec{\phi}_{n}(k_x,y)$ which permit replacing the operator $p_x$ by the quantum number $\hbar k_x$ in Eq. (\ref{HAMIL-KP}). 
We  define the dimensionless canonical operators: 
\begin{equation}
Q(k_x)\equiv \left( \frac{y}{l_B}-k_x l_B \right),
\label{q}
\end{equation}
and
\begin{equation}
P\equiv 
\frac{l_B}{\hbar} p_y,
\end{equation}
combined with the intrinsic energy scale associated to the Fermi velocity,
\begin{equation}
\frac{ \hbar \omega_0}{2}
\equiv  \frac{\hbar v_F}{l_B}.
\end{equation}

Notice that the role of $k_x$ is to shift the eigenvalues of the  $Q(k_x)$ operator.
  It is very convenient to  define the ladder operators: 
\begin{equation}
\alpha(k_x)=  \frac{1}{\sqrt{2}}\left(Q(k_x) + i  P \right),
\end{equation}
which  satisfy
%
$[\alpha(k_x),\alpha(k_x)^{\dagger}]=1$. For simplicity,  in the following we omit the $k_x$ dependence of the $\alpha$ operator.    
We thus can write the Hamiltonian for the $\tau=+1$ valley as:
\begin{eqnarray}
 H_{+}(k_x,p_y)=
\left(
\begin{array}{cc}
\Delta & \frac{-\hbar \omega_0}{\sqrt{2}}\alpha^{\dagger} \\
\frac{-\hbar \omega_0}{\sqrt{2}} \alpha  & -\Delta \\
\end{array}
\right ),
\end{eqnarray}
whereas for the $\tau=-1$  valley the Hamiltonian reads $H_{-}(k_x,p_y)= H(k_x,p_y)^{\dagger}$.

In order to find the eigenfunctions and eigenvalues of these Hamiltonians, it is convenient to compute their square:
\begin{equation}
 H_{\tau=\pm 1} ^2=
\left (
\begin{array}{cc}
{\cal H}^2(\tau) & 0 \\
 0 & {\cal H}^2(-\tau) 
\end{array}
\right),
\label{H2}
\end{equation}
where $ {\cal H}^2(\tau) \equiv \Delta^2 + \frac{1}{2}\left(\hbar \omega_0\right)^2 \left( \alpha^{\dagger}\alpha+\frac{1-\tau}{2}\right) $.

In the following we denote the  the eigenstates of the operator $\alpha^{\dagger}\alpha$ as $\phi_n$, with eigenvalues $n=0,1,...$ and eigenfunctions $\phi_n$.  The states  $\phi_n$ are the standard harmonic oscillator wave functions, centered around $y= k_x l_B^2$.  The eigenstates of  $H_{\tau=\pm 1}^2$, and thereby eigenstates of ${\cal H}_{\tau}$, denoted by $\vec{\phi}_{n}$,  fall in two categories: the so called, zero Landau level,  with a sub-lattice polarized wave function, and the normal Landau levels.  

\subsubsection{ Landau Levels}
From Eq. (\ref{H2}) it can be seen right away that eigenstates can be written as: 
\begin{equation}
\vec{\phi}_{n}(\tau)=\left(\begin{array}{c} A_n \phi_n \\ B_n \phi_{n-\tau} \end{array}\right),
\label{wave-func}
\end{equation}
where $A_n$ and $B_n$ are coefficients that are determined by requesting that $\vec{\phi}_n$ are also eigenstates of the Dirac equation.  The corresponding eigenenergies are $E_n^2= \Delta^2 + \frac{1}{2}\left(\hbar \omega_0\right)^2 \left(n+\frac{1-\tau}{2}\right)$.  Therefore, the general equation for the eigenvalues of the Dirac Hamiltonian under the influence of a perpendicular two dimensional field are: \cite{Koshino-Ando,Vasilopoulos2012}
\begin{equation}
E_n(\tau)= \pm \sqrt{\Delta^2 + \frac{1}{2}\left(\hbar \omega_0\right)^2 \left( n+\frac{1-\tau}{2}\right)}.
\label{energies}
\end{equation}

It is apparent that the Landau level energies  are independent of $k_x$. Therefore, they give rise to flat bands with a very large degeneracy.  
Moreover, there is  an additional twofold  valley degeneracy given by: 
\begin{equation}
E_n(\tau=-1)=  E_{n+1}(\tau=+1),
\end{equation}

Of course, Eq. (\ref{wave-func}) is only mathematically defined if both $n$ and $n+\tau$ are non negative.
As a results,  the minimal value that $n$ can take is $n=1$ for $\tau=-1$, and $n=0$ for $\tau=+1$.   For these states, the energy can be written as 
$E_{N}= \pm \sqrt{\Delta^2 + \frac{1}{2}\left(\hbar \omega_0\right)^2 (N+ 1)}$, where 
$N+1$ is a strictly positive integer.  
In summary, these states come in doublets, on account of the valley degree of freedom, and in addition have  electron-hole symmetry. 

\subsubsection{Zero Landau Level}
In addition to these states,  for each valley there is an extra eigenstate of  
 $H^2$ with the minimal eigenvalue $H_{\tau}^2\vec{z}_{\tau}=\Delta^2\vec{z}_{\tau}$. They  are: 
\begin{equation}
\vec{z}_{\tau=+1}=\left(\begin{array}{c} \phi_0 \\ 0 \end{array}\right),
\end{equation}
and
\begin{equation}
\vec{z}_{\tau=-1}=\left(\begin{array}{c} 0 \\ \phi_0 \end{array}\right).
\end{equation}
We thus see that these wave functions are very special: they are sublattice polarized.  It can be verified right away that these wave functions satisfy:
\begin{equation}
H_{\tau}\vec{z}_{\tau}=\tau \Delta\vec{z}_{\tau}.
\end{equation}

Thus, the energy of the zeroth Landau levels becomes valley dependent 
 due to the mass term $\Delta$, as shown in Fig. 2(a). We can relate  this  to the fact that  the mass introduces an orbital magnetic moment with  valley dependent orientation. \cite{Niu2007}

\begin{figure}[t]
\includegraphics[clip,width=0.5\textwidth,angle=0]{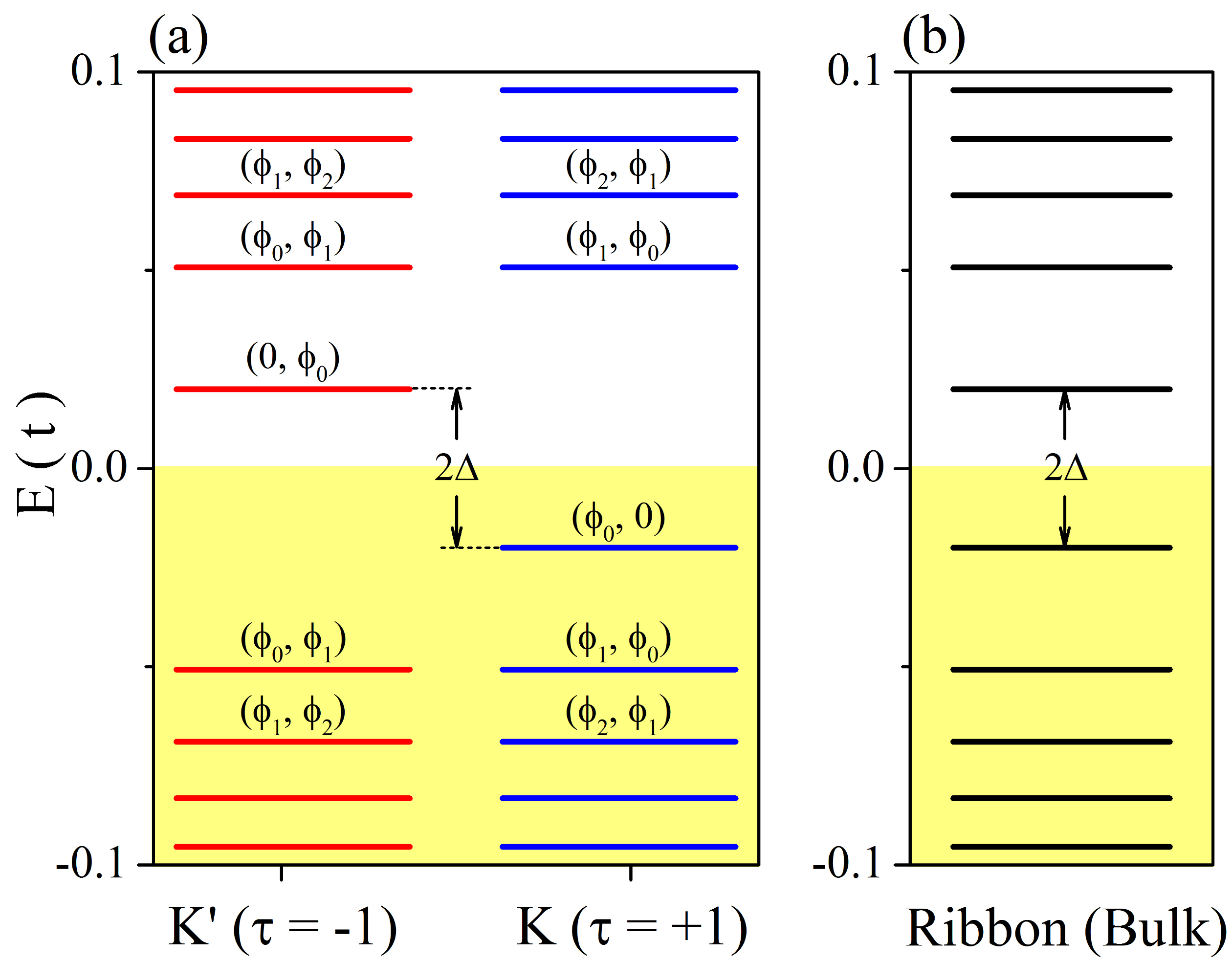} 
\caption{(Color online)
(a) Scheme of the Landau levels for gapped graphene in the $\vec{k}\cdot\vec{p}$ approximation. Notice the valley dependence  for the $n=0$ Landau level.  Labels show the wave function resolved in sub lattice component. (b) Scheme of the Landau levels in graphene ribbon, where valley mixing takes place. The energies of the levels correspond to $\Delta = 0.02$ t and $\Phi=0.0002$ ($B = 15.8$ T).}
\label{Fig2:scheme}
\label{Jhon}
 \end{figure}

A summary of the energy spectrum for gapped graphene electrons under the influence 
of a  perpendicular magnetic field,  described within the $\vec{k}\cdot\vec{p}$ 
 approximation is shown in Fig. \ref{Fig2:scheme}.   Whereas for all the finite 
$|n|$ levels each valley contributes with one Landau level, so that they come in 
couples in graphene,  the $n=0$ Landau levels are valley polarized, so that there 
is only one for the electron sector and one for the hole sector. For $\Delta=0$ 
these two $n=0$  levels are degenerate. However, this degeneracy is lifted for 
gapped graphene, and a gap between them is open.   This discussion has ignored the 
spin degree of freedom, which would add an additional twofold degeneracy to all 
the levels, broken by the Zeeman splitting. 

\subsection{Landau levels and edge states}
We now consider the spectrum of the edge states of  gapped graphene in the 
quantum Hall regime. Edge states are important in this regime because they 
provide the only transport channel.  Whereas it is possible to provide an 
approximate description for edge states within the effective mass 
$\vec{k}\cdot\vec{p}$ approximation used for the bulk states in the 
previous section,  here we apply the tight-binding 
methodology to compute the  the energy levels of graphene stripes of 
width $W$ under the influence of a strong magnetic 
field.\cite{Wakabayashi2001,Brey-Fertig06}

The use of the gauge choice of Eq. (\ref{gauge}),  permits studing quantum Hall bars that are infinite along the $x$ direction and have a finite width along the $y$ axis.  For  a unit cell with $N$ atoms, we obtain  $N$ bands $\epsilon_n(k_x)$. 
We can consider  two geometries,  with either zigzag or armchair edges (see Fig. \ref{Fig1:band}(c-d)).\cite{Nakada96}   
In the rest of the paper we describe the magnetic field in terms of the magnetic 
flux per hexagon in the honeycomb lattice,  $\Phi=\frac{  3 \sqrt{3} Ba_{CC}^2}{2 \Phi_0}$  normalized
to the magnetic flux quantum $\Phi_0=\frac{h}{e}$.
For reference, a normalized magnetic flux of $\Phi = 10^{-4}$ corresponds to $B=7.9$ T.

\begin{figure}[t]
\includegraphics[clip,width=0.5\textwidth,angle=0]{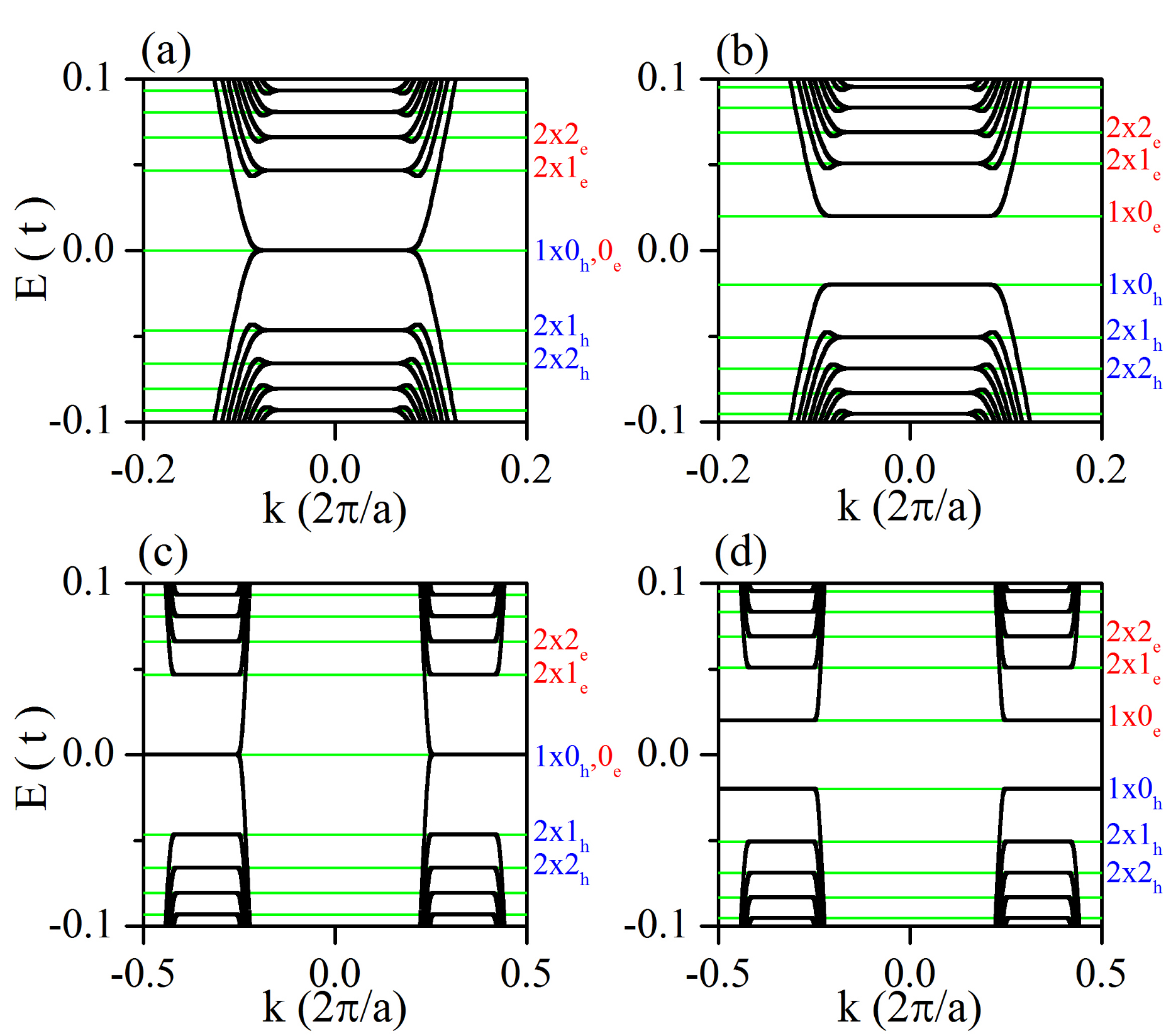} 
\caption{(Color online) 
Band structure in the quantum Hall regime for uniform massless armchair (a) 
and zigzag (c) nanoribbons. 
And of the massive ($\Delta=0.02$ t) ribbon armchair (b) and zigzag (c).
In the four cases the magnetic field is fixed $\Phi=0.0002$ ($B= 15.8$ T) and the dimension are $N_W=1000$ ($W=123$ nm for armchair $W=213$ nm for the zigzag). Green lines correspond to the analytical eigenvalues in Eq. (\ref{energies}).
}
\label{banduni} \end{figure}

In Fig. \ref{banduni} we show the energy bands, denoted by $\epsilon_n(k)$,  
for two different graphene stripes, with zigzag and armchair terminations, for 
$\Phi = 2\times10^{-4}$ and $\Delta=0$ (Fig. \ref{banduni}(a) and (c)) 
and $\Delta=0.02$ t (Fig. \ref{banduni}(b) and (d)). 
The width of the ribbons is $W = 213$ nm for the zigzag and $W =123$ nm
 for the armchair. 
There are several  things to notice. Bands are flat in wide regions of the Brillouin zone and dispersive otherwise. An analysis of the corresponding wave-functions indicate that, except in one case described below, flat bands correspond to Landau level states localized away from the edges. We have verified that the energies are described by Eq. (\ref{energies}).
  In particular, the energy gap between different Landau levels is index dependent, as expected for Dirac electrons and different from non-relativistic electrons. 

 The dispersive states correspond to states localized at the edges.   There is a linear relation between localization along the transverse direction of the ribbon and the momentum $k_x$, as expected from Eq. (\ref{q}). The edge   velocity 
 $v_n(k)=\frac{1}{\hbar}\frac{\partial \epsilon_{n}(k)}{\partial k} $  changes from one edge to the other.
  The emergence  of these chiral edge states whose energy lies in the gap between Landau levels anticipates the very peculiar quantized transport properties of the system, characteristic of the quantum Hall effect.\cite{Halperin82} 

In agreement with the effective mass results,  the flat  bands (Landau levels) have a twofold degeneracy, both in the armchair and zigzag cases,  except for the $n=0$ level.  In the case of armchair termination the degeneracy occurs at the same $k$ point, whereas in the case of zigzag, there are two sets of bands, that can be ascribed to the two valleys.\cite{Brey-Fertig06}  The   $\Delta$ term shifts the position of all the Landau levels and splits the $n=0$ levels opening a transport gap even at the edges,  also in agreement with the effective mass results.  
 The presence of a two flat bands, at a given valley, associated to the $n=0$ Landau is in apparent conflict with the effective mass approximation (see Fig. \ref{Jhon}).
 It turns out that one of the two $n=0$ flat bands at each valley, is an edge state, rather than a bulk state.\cite{Brey-Fertig06}  This statement is further clarified in the next section. 

\begin{figure}[ht!]
\includegraphics[clip,width=0.5\textwidth,angle=0]{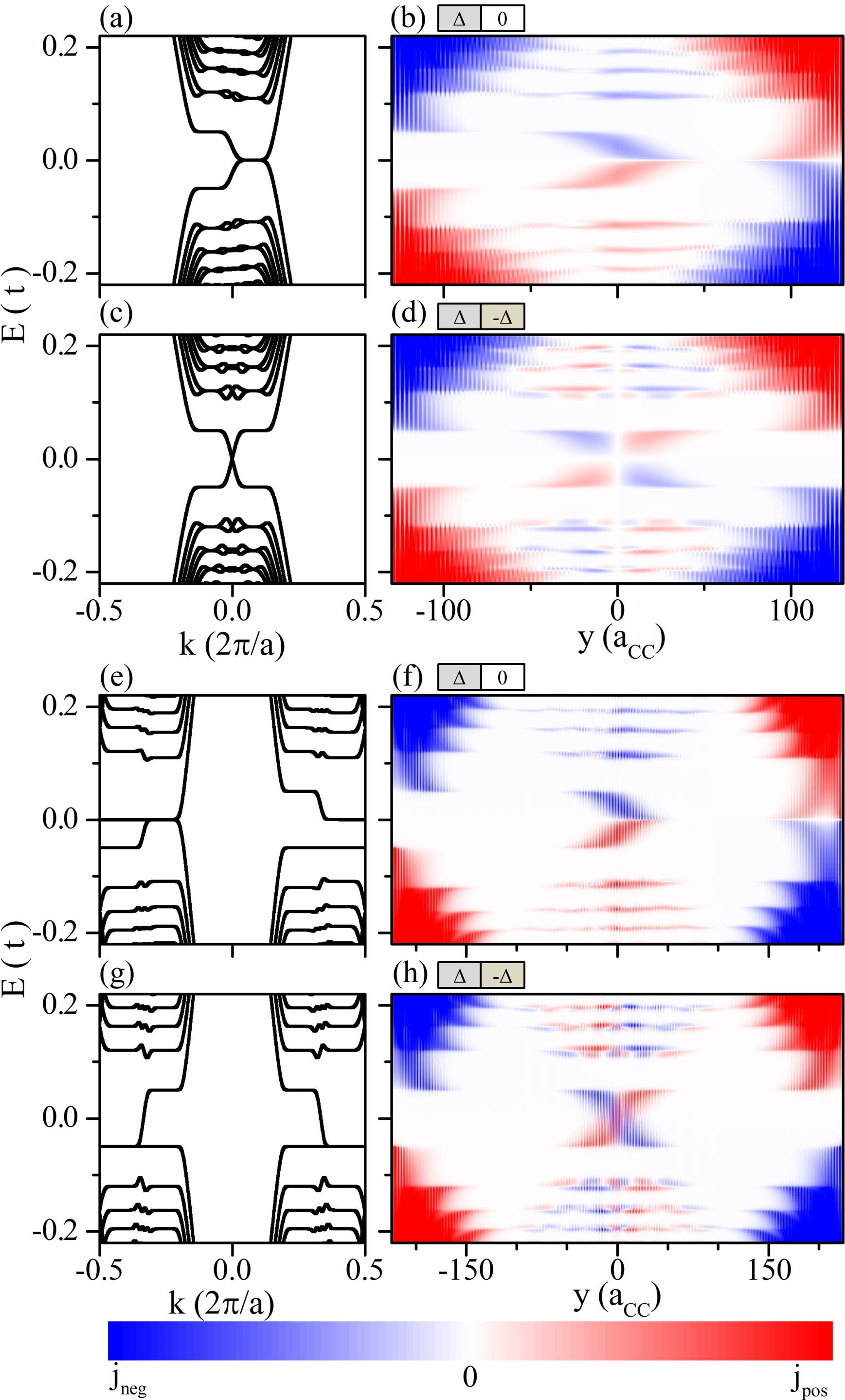} 
\caption{(Color online)
Band structure (left row) and the corresponding  velocity density  
defined by Eq. (\ref{current-density}) (right row). Here we consider transport parallel 
to the interface (see Fig. \ref{Fig1:band} (b)) for armchair (a-d) and (e-h) 
zigzag ribbons.
We fix the mass of the upper half of the ribbon $\Delta_T=0.05$ t, 
and the lower half of the ribbon has either $\Delta_B=0$ ((a-b) and (e-f)) or $\Delta_B=-\Delta_T$ (panels (c-d) and (g-h)). 
The magnetic flux is $\Phi=1.1 \times 10^{-3}$ ($B = 87$ T) 
and the width of the ribbon is $N_W=300$ ($W=36.9$ nm for armchair $W=63.9$ nm for zigzag)  }
\label{bandkink} 
\end{figure}

\section{Electronic properties of graphene quantum Hall bars with inhomogeneous $\Delta$}
We are now in position to  study the electronic states of graphene quantum Hall bars where
$\Delta$ is not constant. For that matter, we consider the simplest situation, a ribbon of width $W$ where  top and bottom halves have  a different mass 
$\Delta_T$ and $\Delta_B$.  We consider two cases $\Delta_T=-\Delta_B$  and $\Delta_T\neq 0,\Delta_B=0$,  both armchair and zigzag terminations (four cases in total).    The band structures, together with the  velocity density:
\begin{equation}
j_x(y,E) \equiv \sum_{k_x,n} |\phi_{n,k_x}(y)|^2\frac{\partial \epsilon_n(k_x)}{\partial k_x}\delta\left(E-\epsilon_n(k_x)\right),
\label{current-density}
\end{equation}
are shown in Fig. \ref{bandkink}.

 For a given Landau level $n$ with wave function  $\psi_n(k,y)$, and within a given valley,  there is a relation between the $k$ quantum number and the average vertical position $\langle  \psi_n(k,y)|y| \psi_n(k,y)\rangle$.  Thus, plots of the velocity density as a function of $y$ provide complementary information to bands $\epsilon_n(k)$.
In panels (a-b) we show the armchair ribbon with  $\Delta_T\neq 0,\Delta_B=0$.  It is apparent that, according to their location in the ribbon,  we can distinguish three types of states: edge states,  bulk states and, in contrast with ribbons with homogeneous $\Delta$,  
interface states  located at the boundary between the massive and massless sectors.  For the bulk states we obtain two different set of flat Landau levels, corresponding to the massless and massive halves, respectively.  The edge states are quite similar to those of the homogeneous mass case.  For the $n\neq 0$ Landau levels, the interface states can be interpreted as  the hybridization of the two pairs of  counterpropagating edge states from the homogeneous $\Delta$ sectors.  
This hybridization results in  two intertwined oscillating bands.  For $n=0$ Landau levels, of the interface states are unique and join the two pairs of $n=0$ Landau levels.   
They can also be interpreted as regular edge states of the massless half confined by the gap on one side and the vacuum on the other.

Results become more interesting for the armchair ribbon with  $\Delta_T=-\Delta_B\neq0$, shown in Fig. \ref{bandkink}(c-d).  Bulk  Landau levels,  edge states and $n\neq0$ interface states are very similar to the previous case. The main difference occurs for the interface states for the $n=0$, which fill the gap almost completely.   The two counter-propagating interface states undergo a small anti-crossing at zero energy.
  These interface states that reside in the gap are quite similar to the kink states reported for bilayer graphene  with a  position dependent off-plane electric field.\cite{Martin-Blanter-Morpurgo, Li-MBM,Alvaro-APL,JeilJung2011}

The discussion for zigzag ribbons goes along the same line. For a  zigzag ribbon with  $\Delta_T\neq 0,\Delta_B=0$. (Fig. \ref{bandkink}(c) and Fig. \ref{bandkink}(f)) we have two replicas of the Landau levels and their edge/interface states for each valley.   For the $n\neq 0$ Landau levels there are bulk flat bands, and dispersive edge and interface states,  very much like in the case of armchair ribbon.  
For the $n=0$ Landau levels,  Fig. \ref{bandkink} shows  four type of bulk states, attending to the sub-lattice ($\sigma_z$) and valley ($\tau_z$)  indexes:
 {\it i)} gapless  with ($\sigma_z=+1,\tau_z=+1$), 
{\it ii)} gapless with ($\sigma_z=-1, \tau_z=-1$), both with zero energy,  
{\it iii)} and {\it iv)} gapped ($\Delta>0$), with energy $\tau_z \Delta$ and either  ($\sigma_z=+1,\tau_z=+1$) or ($\sigma_z=-1, \tau_z=-1$), as expected from the effective mass theory. In addition,  at each valley there is a kink state that joins the gapless $n=0$ Landau level, with the corresponding gapped state.  This kink states  share spectral range with  edge states.

The properties of the zigzag ribbon  with $\Delta_T=-\Delta_B$  are in line with the other cases.  The main feature here is the presence of a kink state at each valley that, in contrast with the armchair ribbon, it has no anti-crossings and covers  the  gap completely. The wave functions of the kink states are located at the interface, as expected.  The velocities of the kink states are opposite for valley.  This is one of the main results of this manuscript: we predict the existence of counter-propagating valley polarized states at the interface of two  graphene quantum Hall bars with opposite masses. 


We have also studied the electronic structure of  ribbons for which  the variation of $\Delta$ is not abrupt (not shown in the figures). For that matter, we have chosen a model with a central region of width $L_{mt}$ where $\Delta$ changes linearly (see Fig. (\ref{aoporl})(d)) .  As long as $L_{mt}$ is smaller than the $l_B$, the bands for this system are qualitatively the same than those shown in 
Fig. \ref{bandkink}.

\section{Transport properties of kink states}

We now discuss the robustness with respect to disorder  of transport properties 
of the kink states found in ribbons with $\Delta_T=-\Delta_B$.  
For that matter, we consider the geometry shown in Fig. \ref{Fig1:band}, an infinite stripe with three regions:  
two semi-infinite electrodes without disorder joined by a central region, 
of length $L$, that features an Anderson disorder potential:\cite{Ziman-book}
\begin{equation}
V=\sum_i V_i c^{\dagger}_i c_i,
\end{equation} 
where $V_i$ is a random variable uniformly distributed over the interval 
$[ -V_0, V_0 ]$, where the energy scale $V_0$ sets the strength of the 
disorder potential. The transmission is calculated for each 
disordered configuration and averaged over a different disorder configuration 
realizations.

Making use of the  partition method and the Green function approach,\cite{Datta-book}
outlined in the appendix \ref{Apen1}, we compute the scattering transmission function $T(E)$,
which relates to the two terminal elastic conductance through the Landauer 
formula $G= \frac{e^2}{h} T(E_F)$, as mentioned  in Eq. (\ref{LandauerG}). 
The transmission function is the sum over of the transmission coefficients $T_n$ 
of the channels $n$ available at a given energy.  For an ideal transmission 
channel without backscattering, $T_n=1$.  A completely blocked channel 
gives $T_n=0$.   

\begin{figure}[t!]
\includegraphics[clip,width=0.5\textwidth,angle=0]{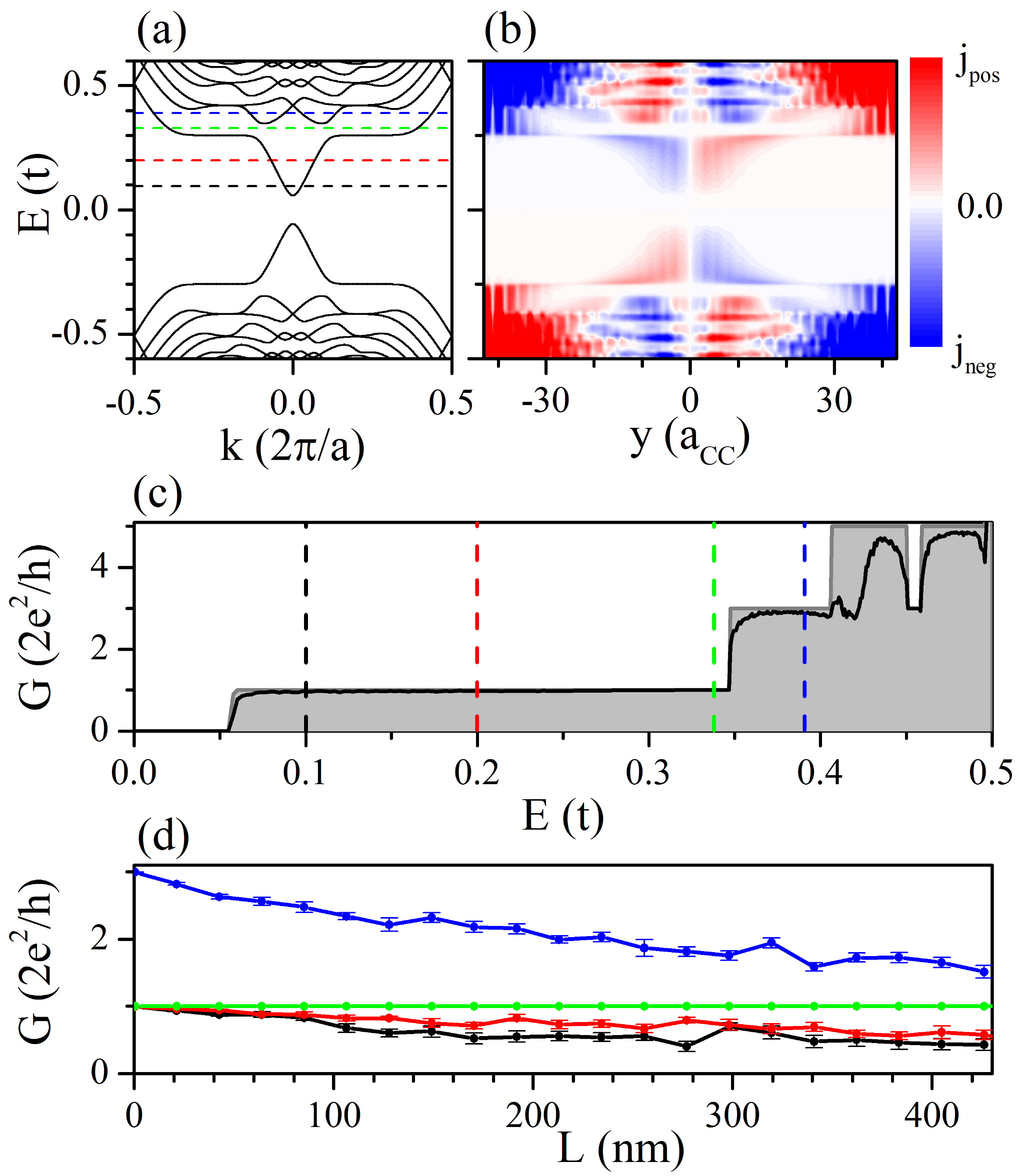} 
\caption{(Color online)
Electronic properties  of an armchair ribbon with a stepwise constant gap $\Delta_T=-\Delta_B=0.3$ t (see Fig. \ref{Fig1:band} (b)), under a magnetic field
$\Phi=0.0081$ ($B =640$ T), $N_W=100$ ($W=12.3$ nm).  
(a) Energy bands, (b) Velocity density map for the disorder-free structure, 
(c) Two terminal conductance as a function of $E_F$ (in units of the hopping t) 
with disorder (Black line). As reference the conductance of the 
disorder-free ribbon has been included  as a shadow region.
(d) Two terminal conductance as a function of channel length for four different energies (marked in (a) and (c)).
In the disorder cases (c-d) the Anderson parameter is $V_0=0.1$ t, an average over $10$ disorder configurations was performed and the error bars reflect the standard deviation.
}
\label{aopoud}
 \end{figure}

\begin{figure}[t!]
\includegraphics[clip,width=0.5\textwidth,angle=0]{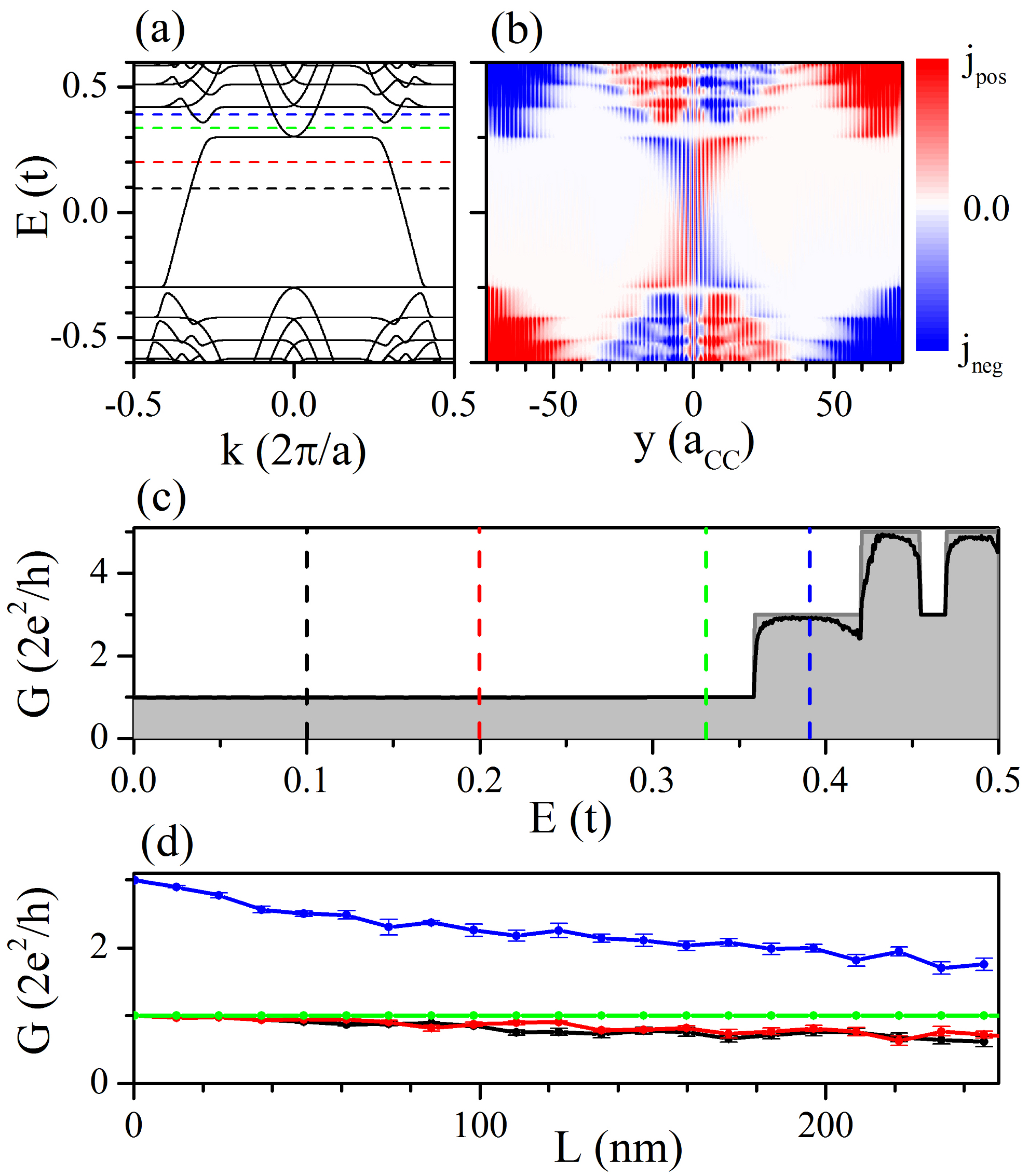} 
\caption{(Color online)
(a), (b), (c) and (d) as in Fig. \ref{aopoud},  for the case of a zigzag ribbon with a stepwise constant gap $\Delta$  with $\Delta_T=-\Delta_B=0.3$ t, under a magnetic field with $\Phi=0.0081$ ($B = 640$ T), $N_W=100$ ($W = 21.3$ nm).  }
\label{zopoud}
 \end{figure}
 
The edge states in quantum Hall bars are the canonical example of ideally 
transmitting channels, with $T_n=1$. This leads to a quantized two terminal conductance,
$G= n \frac{2 e^2}{h}$ , where $n$  is an integer number. The computed  2 terminal  conductance is actually related to the Hall conductance as measured in a 4 terminal Quantum Hall Bar.\cite{Thouless-book}

\begin{figure}[ht!]
\includegraphics[clip,width=0.5\textwidth,angle=0]{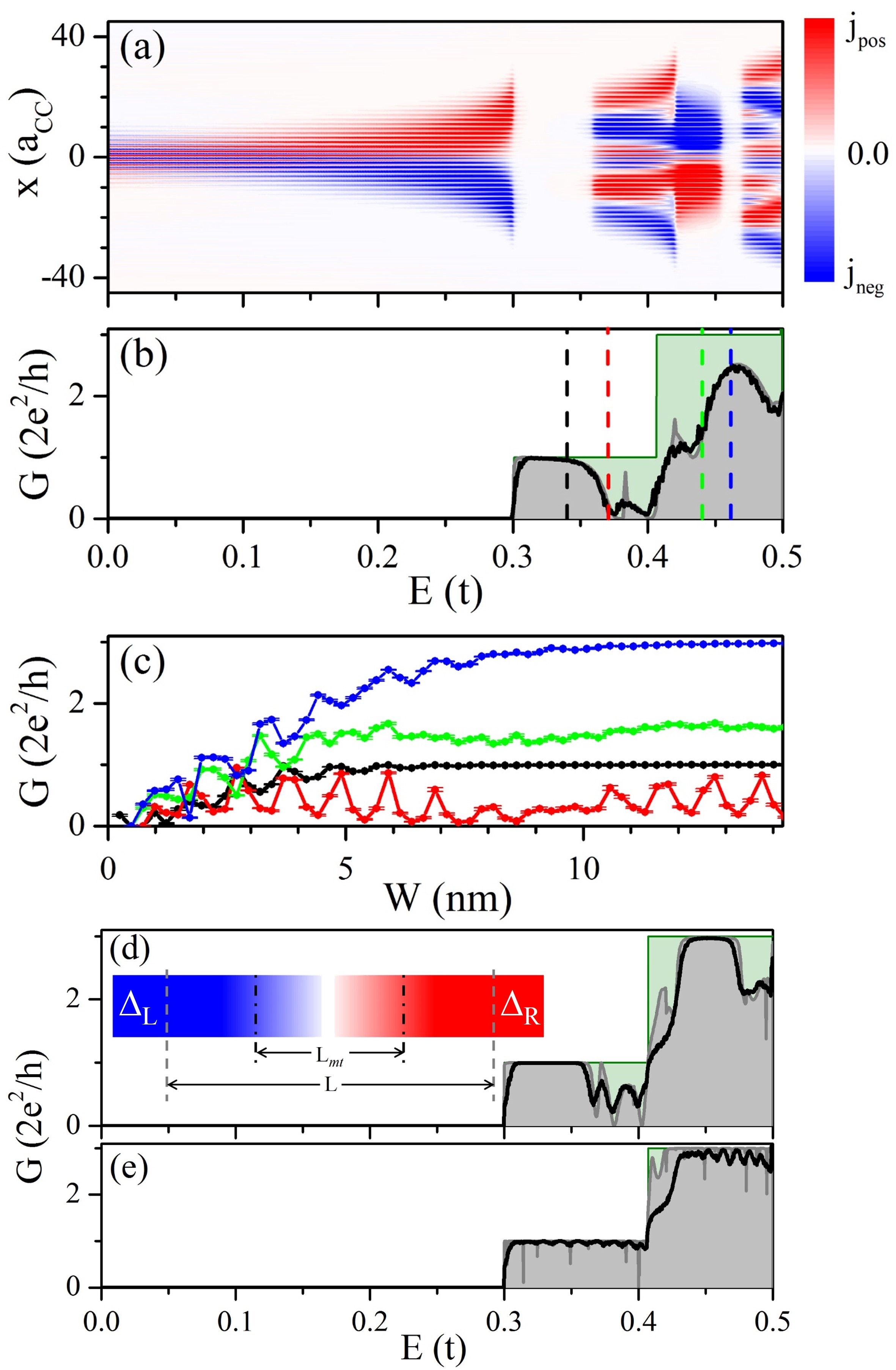} 
\caption{(Color online)
Electronic properties  of a graphene heterostructure made of two semi-infinite  
armchair ribbons with opposite mass $\Delta_L=-\Delta_R=0.3$ t 
(see Fig. \ref{Fig1:band} (a)), under a magnetic field $\Phi=0.0081$ ($B = 640$ T).  
(a) Velocity density map corresponding to the zigzag infinite ribbon 
along the interface.  
(b) Two terminal conductance as a function of $E_F$ with disorder (Black line) for a central region of length $N_L = 30$ ($L = 12.8$ nm) and width $N_W=50$ ($W=6.1$ nm)
and average over 10 disorder configurations was performed. As reference we include the conductance of the disorder-free ribbon has been included as a Grey line for $\Delta_L=-\Delta_R$ and the homogeneous mass $\Delta_L=\Delta_R$ as a Green shadow.
(c) Two terminal conductance as a function of the ribbon width for four different
energies (marked in (b)) calculated with disorder, a channel length $N_L=30$,   averaged over 100 configurations. All the disordered cases have been calculated with Anderson disorder $V_0=0.1$ t.
(d) Conductance as a function of $E_F$, with a linear mass transition form $\Delta_L$ to $\Delta_R$, for a channel length $L = 21.3$ nm and a region of lineal gap transition $L_{mt} = 1.7$ nm in (a) and for $L_{mt} = 8.5$ nm in (d). The color scheme is the same that in (b). %
}
\label{aoporl}
 \end{figure}
 
We consider first the transport properties of the armchair ribbon with
opposite mass $\Delta_T=-\Delta_B$.  Since the calculation of the 
transmission coefficients requires the determination of the Green 
function of a system with $2 \times N_W \times N_L$ atoms, and  we 
consider lengths $N_L$ up to several thousands, it is computationally 
convenient to choose a smaller $N_W$, but large enough so that there is no 
interedge coupling. This also makes it necessary to take unrealistically 
large values of $B$.  However, we expect that the simulated structures  
have the same properties than wider ribbons with smaller magnetic fields.
 
We study now the transmission as a function of energy, for a fixed 
length of the disordered region $N_L$, shown in Fig. \ref{aopoud}(c).
The shadow background with quantized steps shows the transmission without disorder, 
and the black line is the transmission for a length $N_L = 30$ for $V_0 = 0.1$ t. 
The stepwise function for $V_0=0$, i.e., in the absence of disorder,  reflects the number of modes at a given energy,  
starting  from $0$ when the energy lies inside the gap 
(coming from anticrossing of the kink states) and increasing as the Fermi energy reaches new edge states.

\begin{figure}[ht!]
\includegraphics[clip,width=0.5\textwidth,angle=0]{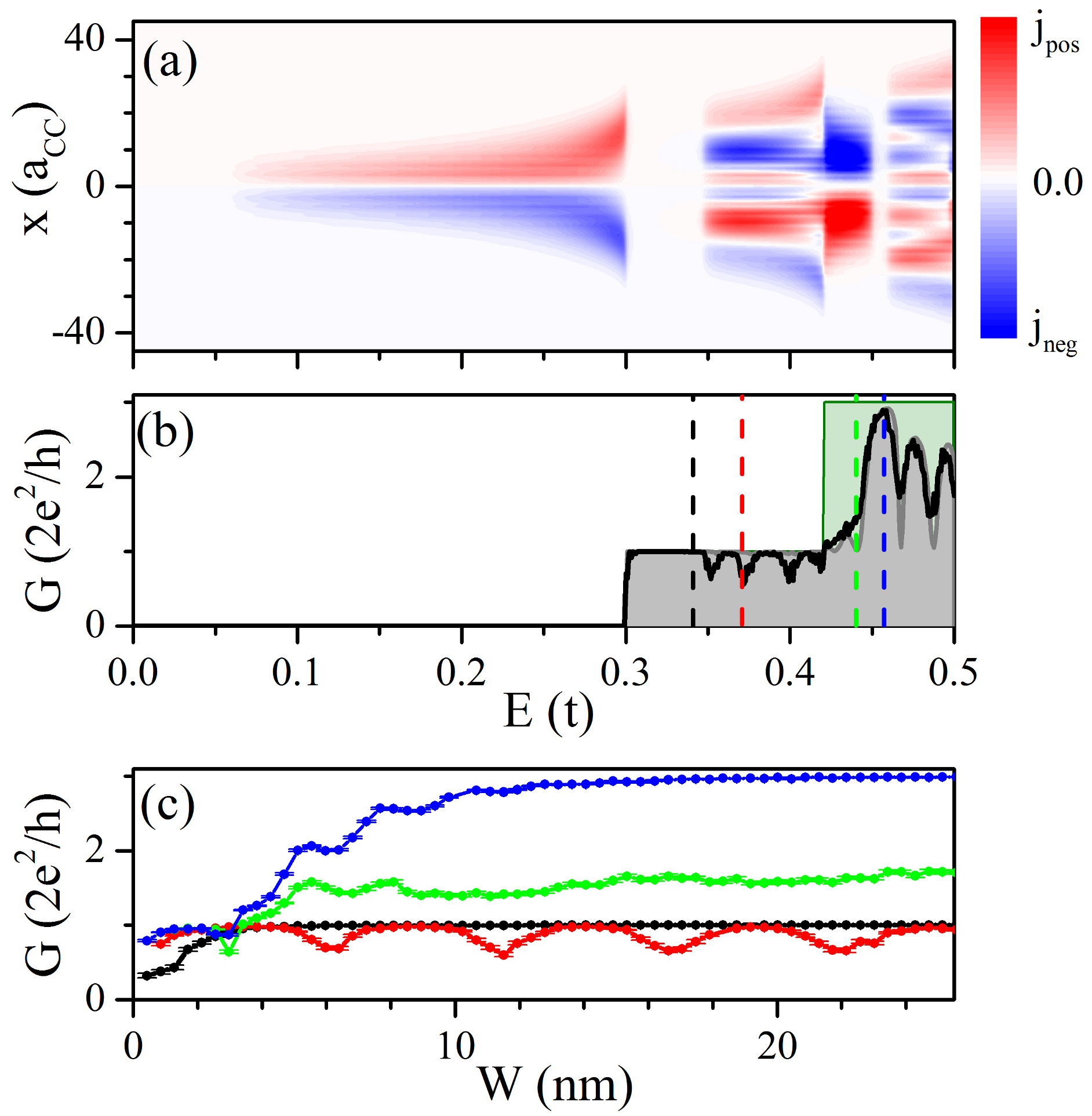} 
\caption{(Color online) (a) Velocity density map corresponding to 
the armchair infinite ribbon  (along the interface, see text). (b) and (c) as 
in Fig. \ref{aoporl}, for the case of zigzag ribbons with opposite mass
$\Delta_L=-\Delta_R=0.3$ t, under a magnetic field with   $\Phi=0.0081$ ($B = 640$ T).  
 }
\label{zoporl}
 \end{figure}

The effect of disorder in Fig. \ref{aopoud}(c), for a fixed channel length, is clearly energy dependent. 
This is more clearly seen in 
  Fig. \ref{aopoud}(d), where  we plot the conductance, averaged over 
10 disorder configurations, at  the four energies marked in Fig. \ref{aopoud}(a, c), 
as a function of the length of the ribbon. The vertical error bars reflect 
the standard deviation.  It is apparent that, 
as the length of the transport channel $L$ increases, the transmission tends 
to the quantized values, when the energy crosses edge states (green line), but 
tends to zero in the case of the kink states (black and red lines).
Furthermore, when the energy crosses both edge and interphase states (blue line) 
the channels corresponding to the interphase states are attenuate as the length 
is increased.  Therefore, backscattering is possible, which is expected since two 
states with opposite velocities coexist in the interface region.

When the same analysis is done for a zigzag ribbon with $\Delta_T=-\Delta_B$, 
the results found are quite similar (see Fig. \ref{zopoud}).  Therefore, our transport results show that, unlike edge states, kink states living in the interface of two gapped graphene regions with opposite gaps are not protected.  It must be stressed that, very much like in the case of gapped bilayer graphene, the existence of two valleys prevents the robustness of kink states with respect to disorder, as it would happen in the case of kink states  located at the domain wall for a  two dimensional electron gas of Dirac electrons.

\section{Quantum Hall transport properties of gapped graphene heterojunctions}

In this section we study transport across of the boundary that separates
 two  gapped graphene Quantum Hall bars. Our main goal is to analyse the backscattering that takes place 
 at the interface due to interface states that connect the edges. 
  A scheme of this   heterostructure,  made of 
two regions with constant mass $\Delta_L=-\Delta_R=\Delta$ that meet at an 
abrupt interface, is shown in   Fig. \ref{Fig1:band}(a). We consider transport across 
the interface region which, in addition to the stepwise constant mass 
has Anderson disorder with $V_0=0.1 $ t over a distance $N_L=30$.
Away from the interface each side of the junction hosts a set of Landau levels 
and edge states as described in the section II.
  Because both electrodes 
have a gap $2 \Delta$, conductance is only possible for states with 
energy $E>\Delta$.  
  
Both sides of the structure host  chiral edge states  
that do not backscatter, even  in the presence of disorder, unless 
electrons  at one edge can undergo scattering to the other edge.  
This could be enabled if interface states, i.e., states running along 
the interface, perpendicular to the transport direction, are available 
at the Fermi energy and  are not completely blocked by  disorder.
Below we show that interface states, perpendicular to the transport direction, 
can act as an efficient shortcut between the right-goers
in one edge and the left-goers in the other 
(see Fig. \ref{Fig1:band}), providing a backscattering channel which destroys the 
conductance quantization. 
It must be stressed that, in this geometry, the in-gap kink states studied 
in the previous section, do not play a role in transport, due to the absence 
of transport states in the electrodes inside the gap.

In Figs. \ref{aoporl} and \ref{zoporl} we show results for the conductance 
of both armchair and zigzag heterojunctions.  The interface between the two armchair 
(zigzag) ribbons with opposite mass is given by a zigzag (armchair) boundary whose
properties can  be related to those of the infinite graphene zigzag (armchair) 
ribbon with inhomogenous mass.  
For that matter, the top panel of Figs.  \ref{aoporl} and \ref{zoporl} shows the velocity density map 
obtained for the infinite ribbon with $\Delta_T=-\Delta_B$. Whereas a priori this velocity density  should not be identical to the boundary of Fig. \ref{Fig1:band}(a), due to the finite width of the  of the Hall bar, our results indicate that it permits to anticipate the existence of backscattering induced by  interface states.  

Since we are interested in the interface states as a source of backscattering, it is not necessary to consider a long central  region with disorder, as we did in the previous section.   In addition, the role of disorder here could be to reduce the efficiency of the interface states to produce  backscattering, improving the conductance thereby. 

For both armchair and zigzag geometries the conductance is zero for $E<\Delta=0.3$ t and is quantized 
in $G=\frac{2e^2}{h}$  above the gap, over an energy interval that coincides 
with the absence of interface states, shown in the panel (a) of  
Figs. \ref{aoporl} and \ref{zoporl}.   
The comparison of the velocity density for an infinite ribbon  with the two terminal 
conductance for both heterostructures,  reveals a relation between the existence of 
interface states in the junction ((a) panels) and the backscattering in transport ((b) 
panels). In particular, the two spectral regions with null velocity density at the interface 
give quantized plateaus of conductance.  This is particularly apparent in the case the  plateau with  $G=\frac{2e^2}{h}$ for $E$ right above the band gap.

The connection between  interface states and backscattering if further 
confirmed by studying the conductance as a function of the width of the Hall bar, 
at four different energies. In Figs. \ref{aoporl}(c) and \ref{zoporl}(c) at 
the energies black and blue,  the backscattering is completely canceled for 
sufficiently wide ribbons. At those energies, there are no interface states. 
In contrast, for energies red and green, the conductance oscillates as a 
function of the ribbon width. 
This can be interpreted as follows. As the ribbon width is increased, 
the discrete spectrum  of interface states shifts. When a interface state 
is in resonance with the electrode states, the backscattering is possible 
and conductance is reduced. In contrast, peaks in the transmission corresponds 
to poor matching between incoming state and interface state.  
It is apparent that the amplitude of the oscillations does not decrease 
significantly as the width of the ribbon increases, even in the presence 
of disorder. This suggest that the localization length of the  interface 
states along the direction perpendicular to transport is longer than the ribbon width.  
The results are qualitatively  similar in the case of a heterostructure made of  
zigzag ribbon with opposite masses. In this case the domain wall separating the 
two regions with opposite mass runs along the armchair direction.  

Disorder seems to have two effects on the interface-state induced backscattering in  these heterojunction.  One one side, is probably increasing the  mixing of edge states to interface states, which should enhance the backscattering. On the other side,  for sufficiently wide ribbons disorder could result in the localization of the interface states that are responsible of backscattering, which should decrease the backscattering.  The comparison of the two curves with and without disorder in the middle panels of Figs. \ref{aoporl} and \ref{zoporl} indicate that, the dominance of one effect over the other depends on  energy.  In general, the interface induced backscattering effect is not qualitatively affected by disorder.

\subsection{Smooth gap transition}
We now briefly discuss the effect on the previous results of a non-abrupt change of the $\Delta$ across the junction.
For that matter, we consider   transport across a region where the gap changes linearly in the direction of the ribbon. 
We assume a device with a fixed total length $L$, which contains a central region determined by $L_{mt}$, where the gap changes from $\Delta_L$ in the left electrode to $\Delta_R=-\Delta_L$ at the right, as depicted in the inset of Fig. \ref{aoporl}(d). Anderson disorder is present in the entire device of length $L$. 
For the sake of briefness we limit our discussion to the case of  armchair ribbons, although  we have also obtained similar results for  the zigzag case.

In Figs. \ref{aoporl}(d,e) we present the two terminal conductance as a function of the energy, for two different values of the length scale $L_{mt}$ that characterize the  soft mass transition, $L_{mt} = 1.7$  and $L_{mt} = 8.5$ nm (or in units of the magnetic length  defined in Eq. (\ref{mag_scale})
$L_{mt} = 0.6\,l_B$ and $L_{mt} = 3$ $l_B$).  It is apparent that for the sharper transition ($L_{mt} = 0.6\,l_B$ ) the curves $G(E)$ (Fig.\ref{aoporl}(d))  is very similar to the  abrupt transition shown in Fig. \ref{aoporl}(b). 
For  the softer mass transition  ($L_{mt} = 3$ $l_B$), shown  in Fig. \ref{aoporl}(e), the backscattering induced at the interface is depleted.

\section{Discussion and conclusions}

We have studied the electronic properties of graphene quantum Hall bars 
with a position dependent mass $\Delta$.  We have considered the case of 
stepwise constant $\Delta$.  We  have found that at the boundary of two 
regions with opposite $\Delta$ both in-gap kink states and interface states appear.  Interface states arise from the mixing of counter-propagating edge 
states that coexist in energy.
In contrast, kink states  arise in the domain wall between two gapped 
regions and do not coexist in energy with bulk states. 
In the case of zigzag ribbons,there is one kink state 
at each valley, whose propagation direction changes from valley to valley.

We have studied  transport in two different  configurations that would permit to probe either the in-gap kink states or the interface states.  The study of transport parallel to the interface between two graphene regions with opposite $\Delta$ in the  quantum Hall regime, permits one to study the properties of the kink states (section III).  We have found that the coexistence in real space of two counter-propagating kink states, corresponding to the two valleys, leaves them unprotected from backscattering created by disorder.    In the case of zigzag ribbon, backscattering requires changing valley, which in turn requires short-range scattering, provided by the Anderson disorder.
 
 The study of transport across the  junction of two semi-infinite ribbons with opposite mass, discussed in section V,  permits studying the effect of interface states as sources of backscattering.  Our calculations show how an incoming electron  to the junction from a chiral edge state could scatter to a kink state propagating from one edge the opposite, enabling backscattering at the specific energies at which interface states exist. 
 
 Our calculations represent a toy model for situations in which graphene quantum Hall bars have a position dependent mass. This could be the case of a mass driven by electronic order, for which different ground states could coexist in the sample, or a mass modulated by the interaction with a substrate with a very large commensuration period.\cite{Mele11,Kindermann2012}

\section*{ACKNOWLEDGMENTS} 
This work has been financially supported by MEC-Spain (Grant Nos. FIS2010-21883-C02-01  
and CONSOLIDER CSD2007-0010,  and Generalitat Valenciana
through Grant Nos. ACCOMP/2012/127 and PROMETEO/2012/011)
J.L.L. would like to gratefully acknowledge to the INL for their hospitality and summer scholarship program.

\appendix
\section{Calculation of Transmission}
\subsection{Partition Method and Green functions} \label{Apen1}

In this appendix we give the technical details of the calculation of the Transmission function $T(E)$,  which yields the two terminal conductance through the Landauer formula.\cite{Datta-book}  The calculation method applies for a one dimensional system that can be split in  three regions,  a central "device" of finite size described with the Hamiltonian matrix $H_c$  coupled to two semi-infinite electrodes, left and right leads, described with $H_L$ and $H_R$.   In matrix form the Hamiltonian reads:
\begin{equation}
  H=
\left(
\begin{array}{cc}
H_C & V \\
V^\dagger & H_{S} \\
\end{array}
\right),
\label{htot}
\end{equation}
where
\begin{equation}
 H_S=
\left(
\begin{array}{cc}
H_L & 0 \\
0 & H_R \\
\end{array}
\right),
\end{equation}
and
\begin{equation}
 V=
\left(
\begin{array}{cc}
V_L & V_R \\
\end{array}
\right) ,
\label{hs}
\end{equation}
with $V_L$ and $V_R$ the coupling to the left and right leads, which we assume to be otherwise decoupled from each other.   A central quantity in the method is the Green function
 $\hat{G}(E \hat{I}-\hat{H})=\hat{I}$, where $\hat{I}$ is the identity matrix.   

The projection of the Green function operator over the central region can be written, after some algebra,  as:
\begin{equation}
 G_C=\left [E \hat{I}- H_C-\Sigma_R-\Sigma_L \right ]^{-1},
\label{greenc}
\end{equation}
where the self energies $\Sigma_{\eta}$ of the $\eta=L,R$ lead are
given by
\begin{equation}
\begin{array}{cc}
\Sigma_{\eta}=V_{\eta} g_{\eta} V_{\eta}^\dagger\\
\end{array},
\end{equation}
and $g_{\eta}=\left(E\hat{I}_{\eta} -\hat{H}_{\eta}\right)^{-1}$ are the projections of the Green function operators over the $\eta=L,R$ spaces.   

The conductance can be calculated in the linear response regime,
within the Landauer formalism as a function of the energy $E$. In
terms of the Green function of the system,\cite{Datta-book} it reads
\begin{equation}\label{LandauerG}
G = \frac{{2e^2 }}{h}T\left( {E } \right) = \frac{{2e^2 }}{h}
{\mathop{\rm Tr}\nolimits} \left[ {\Gamma _L G_C \Gamma _R
G_C^\dagger} \right],
\end{equation}
where  $T\left( {E } \right)$, is the transmission function across
the conductor, and  $\Gamma_{\eta}=i[ {\Sigma _{\eta}  - \Sigma
_{\eta} ^{\dag} }]$ is the coupling between the conductor and the
$\eta=L,R$ lead.

\subsection{Determination of the electrode Green function}
The Eq.(\ref{greenc}) to Eq.(\ref{LandauerG}) are all expressed in terms of the electrode Green function $g_{\eta}$. In particular, when represented in a local basis,  it is the so called surface term of the $g_{\eta}$ matrices which is needed.   When the electrode Hamiltonian is written in the form: 
\begin{equation}
H_{\eta}=
\left (
 \begin{array}{ccccc}
  h_{\eta}&v_{\eta}&0&\cdots&\\
v_{\eta}^\dagger &h_{\eta} &v_{\eta} &0&\\
0&v_{\eta}^\dagger&\ddots&\ddots&\\
\vdots&0&\ddots&\ddots&\ddots\\
 \end{array}
\right),
\label{hlead}
\end{equation}
it can be shown that  the electrode Green function satisfies the self-consistent equation: 
\begin{equation}
 g_{\eta}=\left ( E-h_{\eta}-v_{\eta} g_{\eta} v_{\eta}^\dagger \right )^{-1},
\label{selfcons}
\end{equation}
and corresponds to the central and one of the most time consuming steps in the calculation. 
We label the electrode Green function obtained in step $i$ of the iteration procedure, as $g^i_{\eta}$. We have found that the stability of the self-consistent procedure is improved by using the following algorithm to compute the step $i$, for $i\geq 1$ 
\begin{equation}
 g^i_{\eta}=\alpha g^{i-1}_{\eta}+\beta g^{i-2}_{\eta}+(1-\alpha-\beta)g^{i-3}_{\eta},
 \end{equation}
with the the initial guess $g^{0}_{\eta}=g^{-1}_{\eta}=g^{-2}_{\eta}=g^{-3}_{\eta}$ 
and $\alpha$, $\beta$ are mixing parameters.

\subsection{Calculation of the transmission}

The other source of computational overhead in the calculation of 
Eq. (\ref{LandauerG}) is the inversion of the central region $H_C$ 
matrix, renormalized with the self-energies, to obtain $G_C$. 
However, this can be greatly simplified taking advantage of two facts. 
First, in the computation of the transmission, only a few matrix elements 
of the device Green function are actually needed, in particular, those 
involved in the $\Gamma_{\eta}G_C$ products, which are a minor fraction, 
given the surface nature of the $\Gamma$ matrices.  Second, the device 
Hamiltonian can be written as a tridiagonal block matrix. This permits 
to use specific techniques for  tridiagonal matrices that
make the procedure much faster.\cite{tridia} Taking advantage of this  
approach, it is possible to compute the transmission of $300$ nm  
long bars in a desktop computer.

\end{document}